# Is Private Browsing in Modern Web Browsers Really Private?

Abu Awal Md Shoeb
Department of Computer Science, Rutgers University
abu.shoeb@rutgers.edu

## Abstract

Web browsers are the most common tool to perform various activities over the internet. Along with normal mode, all modern browsers have private browsing mode. The name of the mode varies from browser to browser but the purpose of the private mode remains same in every browser. In normal browsing mode, the browser keeps track of users' activity and related data such as browsing histories, cookies, auto-filled fields, temporary internet files, etc. In private mode, it is said that no information is stored while browsing or all information is destroyed after closing the current private session. However, some researchers have already disproved this claim by performing various tests in most popular browsers. I have also some personal experience where private mode browsing fails to keep all browsing information as private. In this position paper, I take the position against private browsing. By examining various facts, it is proved that the private browsing mode is not really private as it is claimed; it doesn't keep everything private. In following sections, I will present the proof to justify my argument. Along with some other already performed research work, I'll show my personal case studies and experimental data as well.

## Introduction

Web browsers are designed in a way that enables to keep the record and retain a lot of information related to their users' activities. The files such as caching files, visited URLs, search terms, cookies, are stored on the local computer and can be easily accessed and retrieved by any person who uses the same computer. The private browsing mode feature was first introduced in 2005 by Apple Safari 2.0. After three years, it was followed by Google Chrome 1.0 (Incognito). Later, in 2009 Microsoft Internet Explorer 8 and Mozilla Firefox 3.5 introduced their versions of private browsing modes known respectively as InPrivate and Private Browsing [1]. Many papers have been written investigating the privacy mode features provided in modern web browsers including Internet Explorer, Google Chrome, and Firefox, and comparing them to one another. However, in all investigations, it is proved that private browsing is not actually private. Anyone can easily trace the activity of a user performed in private browsing mode [2]. In such investigations, temporary files, main memory, and some system files were considered to disprove the claims of private browsing. However, in my threat model, I mostly considered temporary files maintained by different browsers.

## Threat Model

I categorize attackers same as in Aggarwal et al. [5]. The attackers are into two types; local and remote. A local attacker is someone who has physical access to a user's machine. The primary goal of private browsing is to prevent local attackers. In my experiment, there is no one who grabs my computer right after my private browsing session and gets access to my browsing history. However, my private browsing history is still compromised. So this attack model is considered as a local attack as well. On the other hand, for remote attackers, it is assumed that the attacker is capable to engage with the user in a web browsing session. This session could be done over HTTP(s). This typically happens when a user navigates to a website that is controlled by an attacker. In this case, the goal of the attacker is to detect whether the user is in the private mode. I excluded the threat of remote attackers as because private browsing has never been designed to prevent web tracking [3,11,12,13,14]. However, interested readers can check other privacy-preserving tools such as TOR [15] for the prevention of web tracking.

## Background

I got interested in an analysis of private browsing mode when I had some mixed experiences in the rental car search. We have a group of good friends who always like to travel. As a result, we always go on a road trip by renting cars. In our group, most of the time, I do the job of renting cars, booking hotels, etc. For one such trip, I searched for a rental car in my laptop running on Windows. I used Google Chrome to search a car at www.priceline.com. Unfortunately, I didn't find any good deal at that time. So I stopped searching and closed my browser. Later, I was thinking what would happen if I search for a car on my phone. It may be noted that my phone was connected to a wireless access point. It is the same access point where my laptop was connected as well. Since both of my devices were connected from the same access point, I assumed that the location of both devices would be same on the internet. Well, I searched again for a car at priceline.com using my iPhone. The search result on my phone came as a big surprise to me. On my phone, I found a very good deal for our desired car. Out of my curiosity, I again searched for the same deal on my laptop. The result on the laptop was same as it was before. These two different results confused me and helped me to think how to search it in other ways. Since my mobile is a separate device and brings different results than a laptop, I was thinking how to search anonymously in the same device. Then the private browsing option came to my mind. I thought I could search anything in private browsing by hiding my identity.

Later, I opened an incognito tab in Google Chrome and searched for rental cars on the same website. However, the result was same as normal browsing mode. It clicks me to rethink how much private browsing mode is private?

## The Argument

I use private browsing occasionally. I use both Mozilla Firefox and Google Chrome. I use private mode when I search for special deals for rental cars, hotels, and airline tickets. In my personal experience, I have observed that the private browsing mode doesn't keep my search histories private. I'll be explaining three case studies where I found private mode is false.

**Case Study 1**

There are many discounts available on the rental car website when you search as a new customer. Therefore, I try to search for a car in private mode so that it doesn't know that I'm a returning customer until I login to my account. The security of private browsing strikes in my mind first time when I start using my Facebook account right after searching a rental car deals in private mode. I used incognito mode of Google Chrome to search a rental car at [www.priceline.com](www.priceline.com). Later I closed the incognito mode and start using Google Chrome in normal mode. Then I login to my Facebook account and I was so surprised to see rental car deals coming as Facebook advertisements. I was also surprised to see the ads coming from exactly priceline.com. I have a similar experience in searching hotels and airline tickets.

**Case Study 2**

In case study 1, it might happen that I searched for rental cars in normal browsing mode. Later, I may search for the same deals in private browsing mode. Thus the advertisement from priceline.com appeared on my Facebook. In order to justify this case, I again searched for rental cars on other popular websites such as www.hotwire.com, www.kayak.com, and so on. I searched for rental cars for the location of Boston, Massachusetts. After searching several cars in private browsing mode, I closed all tabs. Later, I opened a new tab in the normal mode of Google Chrome and logged in to my Facebook account. At some point, I started seeing advertisements displayed from Hotwire, Kayak, and Priceline in my Facebook. Later, I browsed other websites such as Google, online newspaper, online shopping site, and found a similar advertisement coming from Hotwire, Kayak, and Priceline. Most interestingly, all rental car deals exactly came for the location of Boston, Massachusetts, what I was exactly searching in private mode.

**Case Study 3**

I have a similar experience in searching tickets for airline and hotels. For example, I searched hotel in New Orleans in private mode and later I received an advertisement for New Orleans hotels on my Facebook, Google pages.

## Research Methodology

My initial plan was to investigate the private browsing mode in all major browsers including Mozilla Firefox, Google Chrome, Internet Explorer, and Apple Safari. I was also planning to run my experiments on the different operating system including Microsoft Windows, iOS, and Ubuntu. However, due to the time and resource limitation, my final experimental data presented is based on Ubuntu. I used only Mozilla Firefox and Google Chrome in Ubuntu. Moreover, I was planning to run different web browsers in separate virtual machines so that they do not get the chance to access temporary internet files. Due to the limitation of memory, I could not run virtual machines on my laptop. As a result, I ended up with running both Firefox and Chrome on a single machine run on the Ubuntu operating system.

For my experiment, I set up a fresh Ubuntu 14.04 LTS on my laptop. It is set up as an independent operating system, not a system that resides inside Windows as s secondary operating system. Once my system is ready, I do not login with Firefox or Chrome to prevent synching my personal settings and browsing history. I did the same thing for my Google account as well, so that it doesn't sync my Google search history as well.

**Case 1**

I opened Google Incognito mode and began to search for buying a new car. I searched for a car from Nissan and Toyota. More specifically, I searched for mini SUVs. Then I closed the browser. After some time I reopened the browser in incognito mode and logged in to my Facebook account. At some point, I start seeing the ads specifically for Toyota RAV4 which is a mini SUV. Figure-1 and Figure-2 show the Facebook page with ads. In Figure 2, it doesn't show an advertisement for Toyota but it shows the ads from www.car.com.

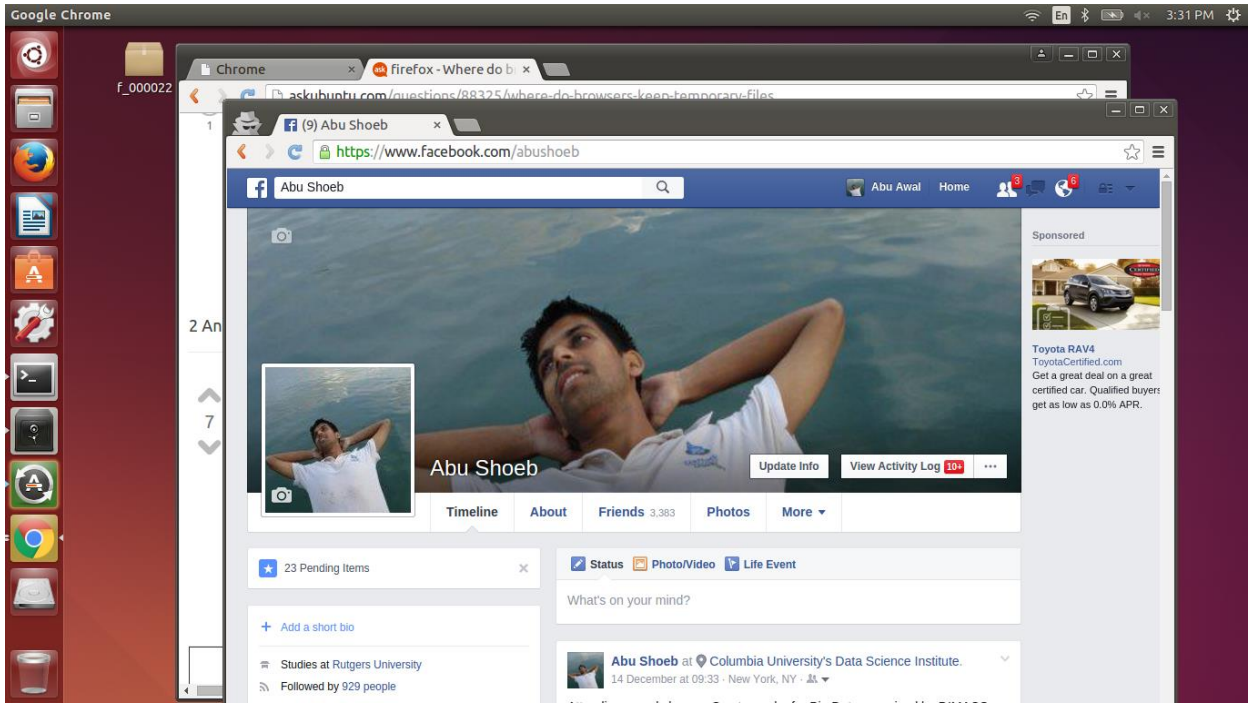

Figure-1

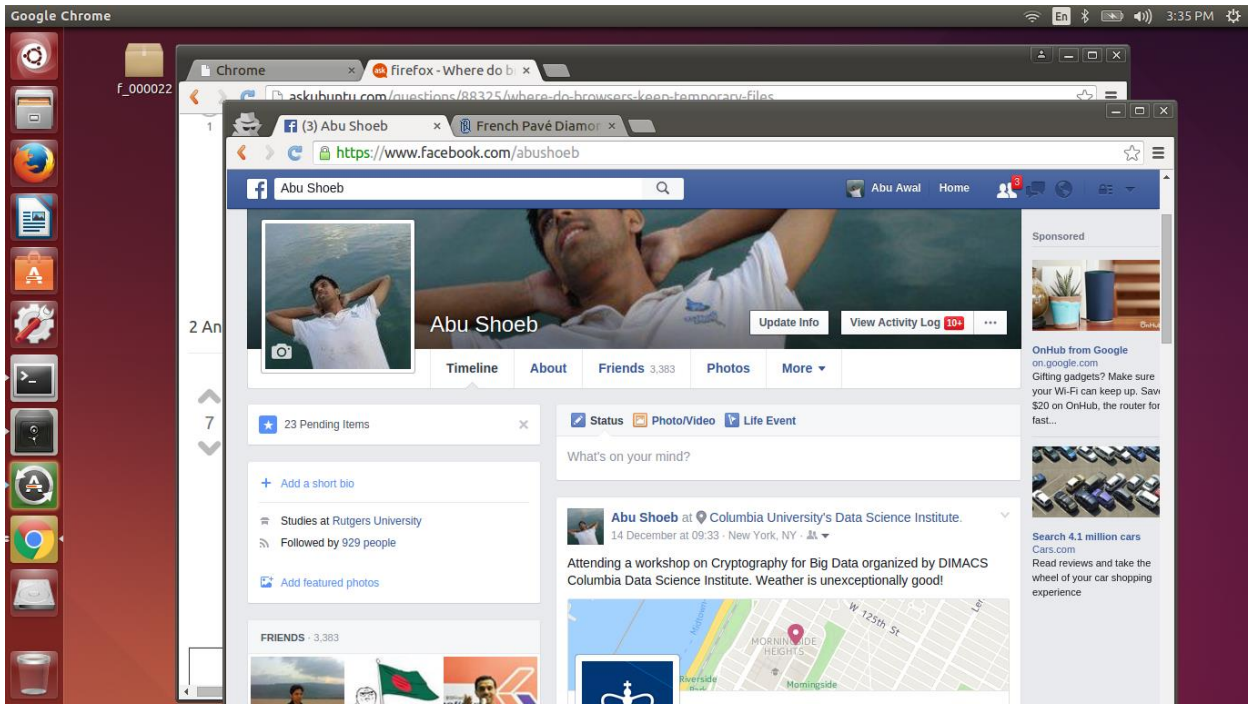

Figure-2

## Case 2

In this experiment, I searched for a completely new item and it was a diamond ring from www.bluenile.com. I used Google Chrome in incognito mode to search the ring. Later I reopened the browser in incognito mode and visited a different website www.ovidhan.org. It is an online Bangla dictionary website to know the meaning of English words in the Bengali language. Once I searched, a word on the website, the result came up with some advertisements and one of them was for diamond rings from www.bluenile.com. Figure-3 shows the advertisements appeared in www.ovidhan.org.

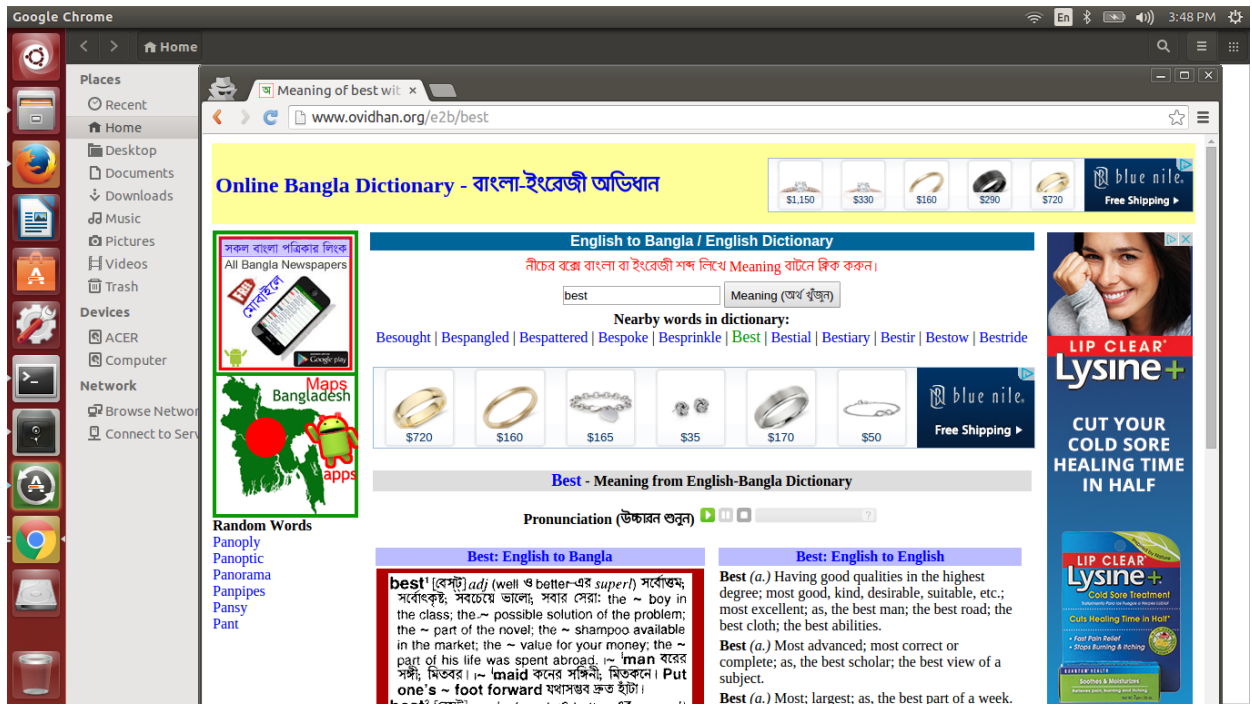

Figure-3

## Related Work

In Aggarwal et al. [5], the attack model includes both site attacker and web attacker. The definition of this attack model is established by the authors. Their study is based on a technique where they discovered how to remotely test if a browser is currently in private browsing mode. They also describe an automated technique to identify failures in private browsing. They discover a few weaknesses in the Firefox browser. Their deepest analysis was conducted in Firefox 3.5.

Howard Chivers's [4] research report tells into the extent that a user's InPrivate browsing history in Internet Explorer 10 can be recovered, and if such records can be reliably identified as resulting from InPrivate browsing. Chivers shows a study based on the InPrivate Browsing on Internet Explorer 10. This version uses a high performance database technology known as the Extensible Storage Engine (ESE) to store data such as internet history and cache memory data. The result of the paper obtained from Windows 8 desktop edition. He shows how the evidence can be found on the increasingly complex data structures used to record the activity on the internet. His findings show where and when the evidence of InPrivate browsing can be retrieved and proved as an artifact private browsing session.

There is another research was performed in Mahendrakar et al. [6] based on some standard tests. They created a website that contained individual pages which required the browser to interact with some forms. They used virtual machine VMWare Workstation 6.5 to perform the tests. They analyzed the existing content in virtual memory after using the browsers Firefox, Internet Explorer, Chrome, and Safari.

Ohana and Shashidhar [7] work with Internet Explorer 8 among other browsers to point out residual artifacts from private and portable web browsing sessions. Portable web browsing artifacts are primarily stored where the installation folder is located (removable disk). They used Microsoft Windows 7 Professional 64 bits as their testbed. They applied their experiment on Microsoft Internet Explorer, Mozilla Firefox, Apple Safari, and Google Chrome.

## Counter Claims

In many organizations, IT department enforces not to do any personal tasks while people work at the office. As a result, they always want to monitor the activity of employees. If a browser provides complete privacy for private browsing then there is a good chance of not using the browser by those offices. Moreover, browsers make revenues from advertising. And an advertisement is shown to users based on usage habits and patterns. If a user surfs in private mode and if the browser doesn't keep any trace of the user then there is less chance of knowing the user's interest. As a result, a customized advertisement can be placed for the user while visiting many pages. By considering these two business cases, browser companies are not willing to provide totally private browsing mode in their browsers.

## Discussion

There is a good number of works have been done to analyze the privacy of private browsing in modern browsers. The study includes both old and new releases of private browsing ranging from 2009 to 2015. The study also covers all modern and popular browsers including Internet Explorer, Mozilla Firefox, Google Chrome, and Apple Safari. Several studies also analyze the content of the private browsing by simulating in virtual machines. Few authors also use computer forensics software to validate the contents of private browsing. In all cases, it was obviously found that private browsing leaves evidence to trace any user activity performed in private mode. The Private mode in Internet Explorer leaves the more evidence than other browsers, then Google Chrome. The Mozilla Firefox keeps less evidence of private browsing. However, there were no browsers found that can completely hide browsing data when it is browsed in private mode.

To summarize, it is found that browser doesn't delete all local data after each session of private browsing. It may delete some files but it doesn't delete everything. I investigated them from command terminal by accessing hidden files maintained by chrome in its cache folder located in /home/user-name/.cache/google-chrome/Default/Cache. Since a browser doesn't have anything in cache folder at the beginning of fresh installation, this could be a potential reason for leaking browsing history. The analysis of the contents of this cache is beyond scope of this paper.

## Conclusion

In this paper, my position is against the private browsing mode for all modern browsers. My claim is that private browsing is not completely private in terms of user's privacy. In this paper, it is proved that private browsing leaves some evidence of users' activity in all cases. These activities can be traced easily by looking into relevant temporary files, cookies, browser's file system, etc. However, the level of privacy varies from browser to browser.

## Future Work

In future, my plan is to test different private browsers on other platforms such as Android. I would like to find out more about the level of the privacy in private browsing in different platforms provided by different companies.

[13] Google Chrome: Browse in private with incognito mode
https://support.google.com/chrome/answer/95464?hl=en

[14] Safari private browsing mode: http://support.apple.com/kb/PH5000

[15] The official website for the TOR project: https://www.torproject.org/